\documentclass[aps,pra,twocolumn]{revtex4-1}
\usepackage{amsmath,amssymb,amsthm}
\usepackage[colorlinks=true,citecolor=blue,urlcolor=blue]{hyperref}
\usepackage{times,txfonts}
\usepackage{braket}
\usepackage{color}
\usepackage{natbib}
\usepackage{amsmath,blkarray}
\usepackage{mathtools}
\usepackage{latexsym}
\usepackage{tabularx, booktabs}
\usepackage{float}
\usepackage{amsfonts}
\usepackage{subcaption}
\usepackage{color,soul}
\begin{document}
\title{Quantum violations of L$\ddot{u}$ders bound Leggett-Garg inequalities for quantum channel}
\author{Asmita Kumari}
\affiliation{Harish-Chandra Research Institute, Allahabad 211091, India}
\author{A. K. Pan}
\email{akp@nitp.ac.in}

\affiliation{National Institute Technology Patna, Ashok Rajhpath, Patna, Bihar 800005, India}
\begin{abstract}
Leggett Garg inequalities (LGIs) provides an elegant way for probing the incompatibility between the notion of macrorealism and quantum mechanics. For unitary dynamics the optimal quantum violation of a LGI is constrained by the L$\ddot{u}$ders bound. In this paper, we have studied two formulations of LGIs in three-time LG scenario, viz., the standard LGIs and  third-order LGIs  both for unbiased and biased measurement settings.  We show that if system evolves under non-unitary quantum channel between two measurements, the quantum violations of both forms of LGIs exceed their respective L$\ddot{u}$ders bounds and can even reach their algebraic maximum in sharp measurement settings. We found that when the measurement is unsharp the quantum violations of both standard and  third-order LGIs for non-unitary quantum channel can be obtained for lower value of the unsharpness parameter compared to the unitary dynamics. We critically examined the violation of L$\ddot{u}$ders bound of LGIs and its relation to the violation of various no-signaling in time conditions, another formalism for testing macrorealism. It is shown that   mere violations of no-signaling conditions are not enough to warrant the violation of standard LGIs, an interplay between the violations of various NSIT condition along with a threshold value  play an important role. On the other hand violation of  third-order LGI  is obtained when the degree of violation of a specific no-signaling in time condition reaches a different threshold value.

\end{abstract}

\maketitle
\section{Introduction}

The notion of macrorealism is a belief which asserts that the properties of macro-objects exist independently at all instant of time and irrespective of our observation. Our everyday world view about the nature of reality is in accordance with that belief. Since the introduction of quantum mechanics (QM), it remains a question as to how such a view can be reconciled with the conceptual framework of QM. Schr$\ddot{o}$dinger \cite{sch} was the first to put forward this question through his famous cat experiment. Since then, this issue has been studied from various perspectives \cite{zur, bruk,ghi}. Motivated by the Bell's theorem \cite{bell64}, in 1985, Leggett and Garg \cite{lg85,lg02} provided a refined definition of macrorealism and formulated a class of inequalities  which provides an elegant route for examining the status of macrorealism in QM. 

Specifically, the notion of macrorealism consists of two main assumptions are the following;
	
\emph{ Macrorealism per se (MRps):} If a macroscopic system has two or more macroscopically distinguishable ontic states available to it, then the system remains in one of those states at all instant of time.
	
\emph{Non-invasive measurement (NIM):} The definite ontic state of the macrosystem is determined without affecting the state itself or its possible subsequent dynamics.

The three-time LG scenario can be encapsulated as follows. Let us consider a dichotomic observable $M_i$ having outcomes $m_i = \pm 1$ at any instant of time $t_i$. Then the classical correlation function between the observables $M_i$ and $M_j$ measured at time $t_i$ and $t_j$ respectively is defined as
\begin{eqnarray}
\label{mij}
 \langle M_i M_j \rangle = \sum_{m_{i},m_{j} = \pm 1} m_{i} m_{j} P(m_{i}, m_{j})
\end{eqnarray}
In three-time LG scenario, the same observable is measured at three different time $t_1, t_2$ and $t_3$ ($t_3 > t_2 > t_1$) which corresponds to the observables $M_{1}, M_{2}$ and $M_{3}$ respectively.  One of the standard LGIs can be written as 
\begin{eqnarray}
\label{lgi1}
L&=& -\langle M_{1} M_{2}\rangle + \langle M_{2} M_{3}\rangle +\langle M_{1} M_{3}\rangle \leq 1
\end{eqnarray}
which is assumed to be valid for a macrorealistic theory satisfying MRps and NIM conditions. By relabeling the measurement outcomes of each $M_i$ as $M_i=-M_i$ with $i=1,2$ and $3$, three more standard LGIs can be obtained.

Standard LGIs are particular class of macrorealistic inequalities. It was argued in \cite{pan18, halli19} that due to sequential nature of the measurement involved in LG scenario, it is more flexible than Bell scenario. Hence,  by keeping the assumptions of MRps and NIM intact there remains possibility to propose various forms of higher order LGIs. The simplest  third-order LGIs is for the three-time measurement scenario. For a dichotomic observable $M$ with outcomes $\pm 1$ sequentially measured at three instant of time $t_1$, $t_2$ and $t_3$, one of the third-order LGIs in three-time  correlation function $\langle M_{1} M_{2} M_3\rangle$, two-time correlation function $\langle  M_{1} M_3 \rangle$ and $\langle  M_{2}\rangle $ is given by  
\begin{eqnarray}
\label{vlgi1}
V=\langle M_{1} M_{2} M_3\rangle + \langle M_{1} M_{3}\rangle -\langle M_{2}\rangle \leq 1
\end{eqnarray}  
Again by relabeling the measurement outcomes of each $M_{i}$ as $M_{i}=-M_{i}$ with $i=1,2$ and $3$, three more third-order LGIs can be obtained. Anther form of third-order LGI is discussed in \cite{halli20}.

 The notion of LGIs have been extensively studied both theoretically \cite{halli20,emary12,clemente15, maroney14, kofler13, budroni15,budroni14,emary,halli16,kofler08,saha15,swati17,pan17,pan20,hal19a,hal21,hal211,sk} and experimentally \cite{lambert,goggin11,knee12,laloy10,george13,knee16,kati1,wang02}. There is a common perception that the LGI is \emph{temporal analog} of Bell's inequalities. This inference is motivated by the structural resemblance between Bell-CHSH inequalities and four-time LGIs. However, LG test shows serious pathology \cite{wild,clemente15,clemente16,halli16,swati17,halli17,pan17,akumari18,pan18,halli19,halli19a} in comparison to a Bell test. The non-signaling in space condition is always satisfied in any physical theory. Thus, the violation of a given Bell's inequality proves a failure of local realism and not the locality alone. In contrast, in the statistical version of the non-invasive measurability assumption, the no-signaling in time (NSIT) can be shown to be violated in QM. Thus, a macrorealist may argue that quantum violation arises due to violation of the NSIT condition. Hence, to draw meaningful conclusions from a LG test, one must satisfy the NSIT condition in QM by adopting a suitable measurement scheme. 

In this work, we restrict ourselves in three-time LG scenario for qubit system.  The optimal quantum value of the standard three-time LGI for a two-level system is $1.5$ (L$\ddot{u}$ders bound). Recently, one of us has proposed that the algebraic maximum of a suitably defined third-order LGIs for a qubit system can approach algebraic maximum by extending the formulation to $n$ time measurement scenario \cite{pan18}. For $n=3$, L$\ddot{u}$ders bound of violation of LGI is $2$ while its macrorealist upper bound is $1$. In \cite{budroni14} it was shown that the quantum violation of LGIs can reach the algebraic maximum of the LG expression for a dichotomic observable of spin-$j$ system qutrit if degeneracy breaking von Neumann rule instead L$\ddot{u}$ders rule is employed. However, the relevance of those results in the context of the LG scenario was criticized in \cite{akumari18}. Recently, by using $PT$-symmetric evolution \cite{bender02}, it was shown that the quantum violation of LGIs in QM can approach the algebraic maximum of three-time four-time \cite{Karthik,varma01,varma} as well as three-time LGIs \cite{javid20}. 

We consider the scenario when instead of unitary dynamics, the system evolves under non-unitary quantum channel between two measurements and examine the quantum violations of standard and third-order LGIs both for biased and unbiased measurement settings.  By considering simplest non-unitary quantum channels - the generalized amplitude damping (GAD) quantum channel we demonstrate that quantum violations of respective L$\ddot{u}$ders bounds of standard and third-order LGIs. Interestingly, the quantum violations of both the forms of LGIs reach their algebraic maximum of three.  We also show that if the measurement is unsharp characterized by positive-operator-valued measure(POVM), then for the unbiased one-parameter family of POVMs, the quantum violations of both standard and third-order LGIs for non-unitary quantum channel can be obtained for a lower value of the unsharpness parameter compared to the case of unitary dynamics.

We note that there are two different ways to perform the LG test. Each of them corresponds to a different form of non-invasive measurability condition, and consequently different notion of macrorealism. As argued by Halliwell \cite{halli16}, a stronger non-invasive measurability condition corresponds to the case when the LG test is carried out through three sequential measurements. In such a case, LGIs are not necessary and sufficient conditions for macrorealism. However, an alternative formulation for macrorealism is proposed by Clemente and Kofler \cite{clemente15} based on the statistical version of the non-invasive measurability condition - the no-signaling in time (NSIT) conditions. They demonstrated that suitable conjunction of two-time and three-time NSIT conditions provide the necessary and sufficient conditions for the stronger notion of macrorealism.  On the other hand, if the LG test is performed through pair-wise sequential measurements, as in the case of the Bell experiment, the non-invasive measurability condition becomes weaker and consequently provides a weaker notion of macrorealism.  It has been argued \cite{halli16} that a suitable set of two-time and three-time LGIs provide necessary and sufficient conditions for the weaker notion of macrorealism. 

We critically examine the violation of L$\ddot{u}$ders bound of LGIs and its relation to the violation of various no-signaling in time conditions corresponding to the stronger notion of macrorealism involving three successive measurements. For the case of standard LGIs, we show that mere violations of no-signaling conditions are not enough to warrant the violation of standard LGIs; it depends on an interplay between the violations of various NSIT conditions and on a relevant threshold value. For the case of third-order LGIs, the violation of third-order LGIs is obtained when the degree of violation of a specific NSIT condition reaches a different threshold value.

This paper is organized as follows. In Sec. II,  we introduce preliminaries corresponding to standard and third-order LGIs, unsharp measurements, and GAD channel operation. In order to make our paper self-contained, we recapitulate the known results for standard and third-order LGIs in sharp and unsharp measurement scenarios while the system evolves under unitary dynamics in Sec.III. In Sec. IV, we study the quantum violation of both the standard and third-order LGIs while the system evolves under the GAD channel and demonstrate how the quantum value of both LG expressions reaches their algebraic maximum. In order to know the reason for getting maximum violation of LGIs for the system state evolved under non-unitary quantum channel in Sec. V we have expressed LG expression in terms of NSIT conditions and shown that how only specific choice of NSIT condition is responsible for the quantum violation of both standard and third-order LGIs beyond L$\ddot{u}$ders bound. Finally, in Sec. VI, we have provided a brief summary and discussion of our results.

\section{Preliminaries}
 Let us briefly encapsulate the notions of the evolutions of qubit system under unitary dynamics and non-unitary quantum channel in three-time LG scenario. 

\subsection{Sequential measurement and evolution of state under unitary dynamics}
Let us consider the system to be two-level quantum state $\rho(t_1)=|\psi(t_1) \rangle \langle \psi(t_1)|$ at $t_1$.
and the measurements are represented as two-parameter family (also called biased) positive operator-valued measures (POVMs) are of the form
\begin{equation}
\label{GO}
M_{i}^{\pm}(\alpha,\vec{n_{i}})=\frac{{\mathbb I} \pm ( \alpha {\mathbb I}+\vec{n_{i}}.\sigma)}{2}
\end{equation}
with $|\alpha|+|\vec{n_{i}}|\leq1$, where $|\alpha|$ is the biasedness and $|\vec{n_{i}}|$ is the sharpness parameter. Here $i=1,2,3$ corresponds to the measurement performed at $t_1$, $t_{2}$ and $t_3$. Clearly, $M_{i}^{+}(\alpha,\vec{n_{i}})+M_{i}^{-}(\alpha,\vec{n_{i}})=\mathbb{I}$. Note that, biased POVMs reduces to the unbiased (spin) POVMs when $\alpha=0$.

The probability of an outcome $m_1=+1$ of the measurement of observable $M_1$ at $t_{1}$ is then given by $tr(\rho(t_1)M_1^{+})$ and the post-measured density matrix can be written as
\begin{eqnarray}
 \rho_{+}(t_{2})=(\sqrt{M_{1}^{+}}\rho(t_1)\sqrt{M_{1}^{+}}^{\dag})/tr(\rho(t_1)M_{1}^{+})
\end{eqnarray}
The joint probability of different outcomes for two POVMs can  then be calculated by using the formula given by
\begin{eqnarray}
\label{pb}
P(m_1, m_2)=Tr(U_{\Delta t_{12}}\sqrt{M_1^{m_1}}\rho(t_1)\sqrt{M_1^{m_1}}^\dag U_{\Delta t_{12}}^\dag M_2^{m_2})
 \end{eqnarray}
where  $m_1, m_2 =\pm 1 $ are the outcomes of measurements at $t_1$ and $t_2$respectively. 

\subsection{Sequential measurement and evolution of state under non-unitary quantum channel}
Instead of unitary dynamics the realistic system evolve under non-unitary quantum channel. Quantum operations of open quantum system are represented by completely positive trace preserving map (CPTP) defined as $\varepsilon (\rho) = \sum_i{K_i \rho K^{\dagger}_i}$. Here $K_i$ is Kraus operator satisfying the completeness relation $\sum_i{K^{\dagger}_i} K_i = \textsl{I}$. CPTP maps is most general operation which maps a valid density matrix to a valid density matrix. In other words CPTP map follows the properties given below:

\emph{1. Linearity:} $\varepsilon (a \rho_1 +b \rho_2) = a \varepsilon (\rho_1) +  b \varepsilon (\rho_2)  $ .	
\emph{2. Preserves Hermiticity: $\rho_1 = \rho_1^{\dagger} \Rightarrow \varepsilon (\rho_1) = \varepsilon (\rho_1)^{\dagger}$}.

\emph{3. Preserves positivity: $\rho_1 \geq 0 \Rightarrow \varepsilon (\rho_1) \geq 0 $}.

\emph{4. Preserves trace: $Tr[\varepsilon (\rho_1)] = Tr[\rho_1]$}.

These CPTP maps are also known as quantum operations, superoperator or non-unitary quantum channels. One of the most important application of non-unitary quantum channel is energy dissipation due to energy loss in quantum system. In order to describe energy dissipation at all temperature GAD channels are used \cite{sri}. For our purpose in this work the consideration of GAD channel is sufficient and the Kraus representation of it is given by
$$G_0= \sqrt{p}\left[ \begin{array}{cc} 1 & 0 \\ 0 & \sqrt{1-\gamma(t)}\end{array} \right], G_1 =\sqrt{p}\left[ \begin{array}{cc} 0 & \sqrt{\gamma(t)} \\ 0 & 0\end{array} \right]$$
$$G_2=\sqrt{1-p}\left[ \begin{array}{cc} \sqrt{1-\gamma(t)} & 0 \\ 0 & 1\end{array} \right], G_3 =\sqrt{1-p}\left[ \begin{array}{cc} 0 & 0 \\ \sqrt{\gamma(t)} & 0\end{array} \right]$$
 where, $\gamma(t)$ is the damping parameter. GAD reduces to amplitude damping (AD) channel either at $p=0$ or at $p=1$.

In order to find the two-time correlation functions of LGIs using GAD channel let us denote the damping parameter of GAD channel as $\gamma_{12}$, $\gamma_{23}$ and $\gamma_{13}$ for the correlation functions in the time interval $t_2 - t_1$, $t_3 - t_2$ and $t_3-t_1$ respectively and corresponding GAD channels for the time interval $t_2 - t_1$, $t_3 - t_2$ and $t_3-t_1$ are denoted by $G_{l}(t_{12})$,  $G_{l}(t_{23})$ and $G_{l}(t_{13})$ ($l = 0,1,2,3$) respectively. If the initial state of the system is $\rho(t_1)$, then the density matrix of the system at time $t_2$ evolves to 
\begin{eqnarray}
\label{evch}
\rho(t_2) & = & G_{0}(t_{12}) \rho(t_1) G^{\dagger}_{0}(t_{12}) + G_{1}(t_{12}) \rho(t_1) G^{\dagger}_{1}(t_{12}) \\ \nonumber &+& G_{2}(t_{12}) \rho(t_1) G^{\dagger}_{2}(t_{12})+G_{3}(t_{12}) \rho(t_1) G^{\dagger}_{3}(t_{12})
\end{eqnarray}
In our work we use Eq. (\ref{evch}) for calculating the evolved state in various stages of our calculation.

\section{Violation of LGI when system evolve under unitary dynamics}
\subsection{Quantum violation of standard LGI}
We first provide the L$\ddot{u}$ders bound when the system evolve under unitary dynamics in between two measurements. The quantum mechanical expression of $L$ (say, $(L_u)_Q$ where $u$ stands for unitary evolution) for a state  $\rho(t_1)$. The various pair-wise correlations in $L$  can be obtained by using joint probability defined in Eq.(\ref{pb}) for the  POVMs in Eq.(\ref{GO}). These are given in  Eq.(\ref{m12}), Eq.(\ref{m23}) and Eq.(\ref{m13}) of Appendix A.

Now, for $\alpha=0$, we obtain $(L_{u})_{Q}$ for unbiased POVMs $M_{1}^{\pm}(0,\eta)=(\mathbb{I}\pm \eta\sigma_{z})/2$, is given by 
\begin{eqnarray}
\label{sl1}
(L_{u})_{Q}=\eta ^2 \left[- \cos (2 g_1)+ \cos (2 g_2) + \cos (2 (g_1 +g_2)) \right]
\end{eqnarray}
which is state independent. The optimal quantum value (L$\ddot{u}$ders bound) of $(L_{u})_{Q}$ is $1.5$ obtained for sharp measurement ($\eta=1$) at $g_1 \approx 2 \pi/3$ and $g_2 \approx \pi/6$. However, violation of standard LGI is obtained if $\eta>0.81$.

By considering a particular form of biased POVM by putting $\alpha = 1- \eta$ in Eq.(\ref{GO}) we have $M_{1}^{\pm}(1-\eta,\eta)=\eta (\mathbb{I} + \sigma_{z})/2$. For this particular choice of POVMs the quantum  mechanical expression of $L$ is given in Eq.(\ref{bsl1}) which is calculated  for the state $\rho(t_1)=|\psi(t_1)\rangle\langle \psi(t_1)|$ where 
\begin{equation}
\label{state}
|\psi(t_1)\rangle = \cos\theta |0 \rangle + e^{i\phi} \sin\theta |1 \rangle
\end{equation}

  From the analysis of Eq.(\ref{bsl1}) we found that in case of biased POVMs, violation of LGI is obtained for any value of biasedness parameter. However maximum violation of $1.5$ is achieved at $\eta = 1$. It is numerically checked that this is the optimal bound for any states. Note that, in case of biased measurement, $(L_{u})_{Q}$ in Eq.(\ref{bsl1}) is state dependent but in case of unbiased measurement $(L_{u})_{Q}$ in Eq.(\ref{sl1}) is state independent. The  L$\ddot{u}$ders bound of standard LGI is $1.5$, irrespective of system size \cite{hal21,budroni13}.

\subsection{Quantum violation of third-order LGI}
 The quantum mechanical expression (say, $(V_u)_Q$) of $V$ in ineq.(\ref{vlgi1}) is calculated for a state in Eq. (\ref{state}). The triple-wise correlation $\langle M_{1} M_{2} M_3\rangle $ in ineq.(\ref{vlgi1}) is defined as
\begin{eqnarray}
\label{pb3}
&&P(m_1, m_2, m_3)\\
\nonumber
&=&Tr(U_{\Delta t_{23}}\sqrt{M_2^{m_2}}U_{\Delta t_{12}}\rho(t'_1) U_{\Delta t_{12}}^\dag\sqrt{M_2^{m_2}}^\dag U_{\Delta t_{23}}M_3^{m_3})
 \end{eqnarray}
with $\rho(t'_1) = \sqrt{M_1^{m_1}}\rho(t_1)\sqrt{M_1^{m_1}}^\dag$ in 
\begin{eqnarray}
\nonumber
\label{mijk}
 \langle M_1 M_2 M_3 \rangle =\sum_{m_{1},m_{2},m_{3} = \pm 1} m_{1} m_{2} m_{3} P(m_1, m_{2}, m_{3}).
\end{eqnarray}
It should be noted here that in three-time LG scenario, the observable $M_{1}$ is measured at three different time $t_1, t_2$ and $t_3$ ($t_3 > t_2 > t_1$) which we denote as the observables $M_{1}, M_{2}$ and $M_{3}$ respectively.\\

The joint, triple-wise and a single expectation value of observable $M_2$ in $(V_{u})_Q$ calculated for the state $|\psi(t_1)\rangle$ in Eq.(\ref{state}) and POVMs  $M_{1}^{\pm}(0,\eta)=(\mathbb{I}\pm \eta\sigma_{z})/2$  are given in  Eq.(\ref{m13}) ,  Eq.(\ref{m123}) and Eq.(\ref{m2}) respectively. 

For the case of third-order LGI, the quantum mechanical expression of $V$ for $\alpha=0$ (unbiased POVM) is obtained as 
\begin{eqnarray}
\label{vsl2}
(V_{u})_Q&&= \eta  \bigg[\cos (2 \theta) \left(\eta ^2 \cos (2 g_2)-\cos (2 g_1)\right)\\ \nonumber &&+\eta  \cos (2 (g_1+g_2))-\sin (2 g_1) \sin (2 \theta) \sin (\phi )\bigg]
\end{eqnarray}
It can be seen that  $(V_{u})_Q > 1$ i.e., violation of third-order LGI is obtained for unbiased POVM if $\eta > 0.62$. The maximum quantum value is obtained to be $2$ for sharp measurement ($\eta=1$) at $g_1 = 3 \pi/4$, $g_2 = \pi/4$, $\theta = \pi/4$ and $\phi = \pi/2$. It is also checked that mixed state provides less violation than pure state. Note that, unlike standard LGI, third-order LGI is state dependent even for the case of sharp measurement. 
By taking a specific form of biased POVM when $\alpha = 1- \eta$, the quantum  mechanical expression $(V_{u})_Q$ is placed in Eq.(\ref{bvl1}). However, this  does not add anything new to the results than that is already known above.

\section{Violation of LGIs : System evolved under non-unitary quantum channel}
\subsection{Quantum violation of standard LGI}
In order to examine the incompatibility between standard LGI and QM for the qubit system evolved under GAD channel, let us again assume that the initial state of the system is same as given in Eq.(\ref{state}) and POVMs given in Eq.(\ref{GO}). It is numerically checked that the results cannot be better for mixed state. In such a case the pair-wise quantum correlations in $L$ are obtained as
\begin{eqnarray}
\label{Gm12}
(\langle M_{1} M_{2}\rangle_{\gamma})_Q&=& \alpha ^2-(\gamma_{12}-1) \eta ^2+\alpha  \gamma_{12} \eta  (2 p-1)\\ \nonumber &+&\eta  \cos (\theta ) [\gamma_{12} \eta  (2 p-1)-\alpha  (\gamma_{12}-2)]
\end{eqnarray}
\begin{eqnarray}
\label{Gm23}
(\langle M_{2} M_{3}\rangle_{\gamma})_Q &=& \alpha ^2+(\gamma_{12}-1) \eta  \cos (\theta ) (\alpha  (\gamma_{23}-2)\\ \nonumber &+&\gamma_{23} \eta  (1-2 p))-\alpha  \eta  (2 p-1) ((\gamma_{12}-1) \gamma_{23}\\ \nonumber&-&2 \gamma_{12})+ \eta ^2 \left(\gamma_{23} \left(\gamma_{12} (1-2 p)^2-1\right)+1\right)
\end{eqnarray}
and
\begin{eqnarray}
\label{Gm13}
\nonumber
(\langle M_{1} M_{3}\rangle_{\gamma})_Q &=& \alpha ^2-(\gamma_{13}-1) \eta ^2+\eta  \cos (\theta ) (\gamma_{13} \eta  (2 p-1)\\ &&-\alpha  (\gamma_{13}-2))+\alpha  \gamma_{13} \eta  (2 p-1)
\end{eqnarray}
 where $\gamma_{12}$, $\gamma_{23}$ and $\gamma_{13}$ are channel parameters.

For $\alpha=0$ (unbiased POVMs), the quantum mechanical expression $(L_{\gamma})_Q$ of standard LGI can be written using Eq.(\ref{Gm12}), Eq.(\ref{Gm23}) and Eq.(\ref{Gm13}) is given by 
\begin{eqnarray}
\nonumber
\label{gsl1}
(L_{\gamma})_Q&= &\eta  \bigg[\gamma_{12} +\eta +\cos (\theta ) \big(\gamma_{12}+\eta ^2 \big(\gamma_{23} (\gamma_{12} (1-2 p)^2\\ \nonumber &+&-1)+1\big)+\gamma_{13} \eta  (2p-1)-1\big)-2 \gamma_{12} p-\gamma_{13} \eta\\ &+&(\gamma_{12}-1) \gamma_{23} \eta ^2 (1-2p)\bigg]
\end{eqnarray}
 The optimal quantum value of $(L_{\gamma})_Q$ is obtained to be $3$ (algebraic maximum) for sharp measurement ($\eta=1$) at $\theta =0,\gamma_{12} = 1$, $\gamma_{23} = \gamma_{13} = 0$ and $p=0$. However, even the quantum violation reaches the algebraic maximum the lower bound of $\eta$ is not sequentially reduced. It is found that the quantum violation of standard LGI is possible if $\eta > 0.58$ compared to $\eta > 0.81$ in unitary case.

Next, for $\alpha=1-\eta$ (biased POVMs), the quantum mechanical expression of standard LGI $(L_{\gamma})_Q$ is given in Eq.(\ref{gsl2}). In this case the quantum violation of standard LGI can be obtained if $\eta > 0.5$. In order to understand the role of channel parameter in the violation of standard LGI the quantum value of $(L_{\gamma})_Q$ against $p$ and $\gamma_{12}$ is plotted in Fig. 1.
\begin{figure}[ht]
\includegraphics[width=9 cm,height=7 cm]{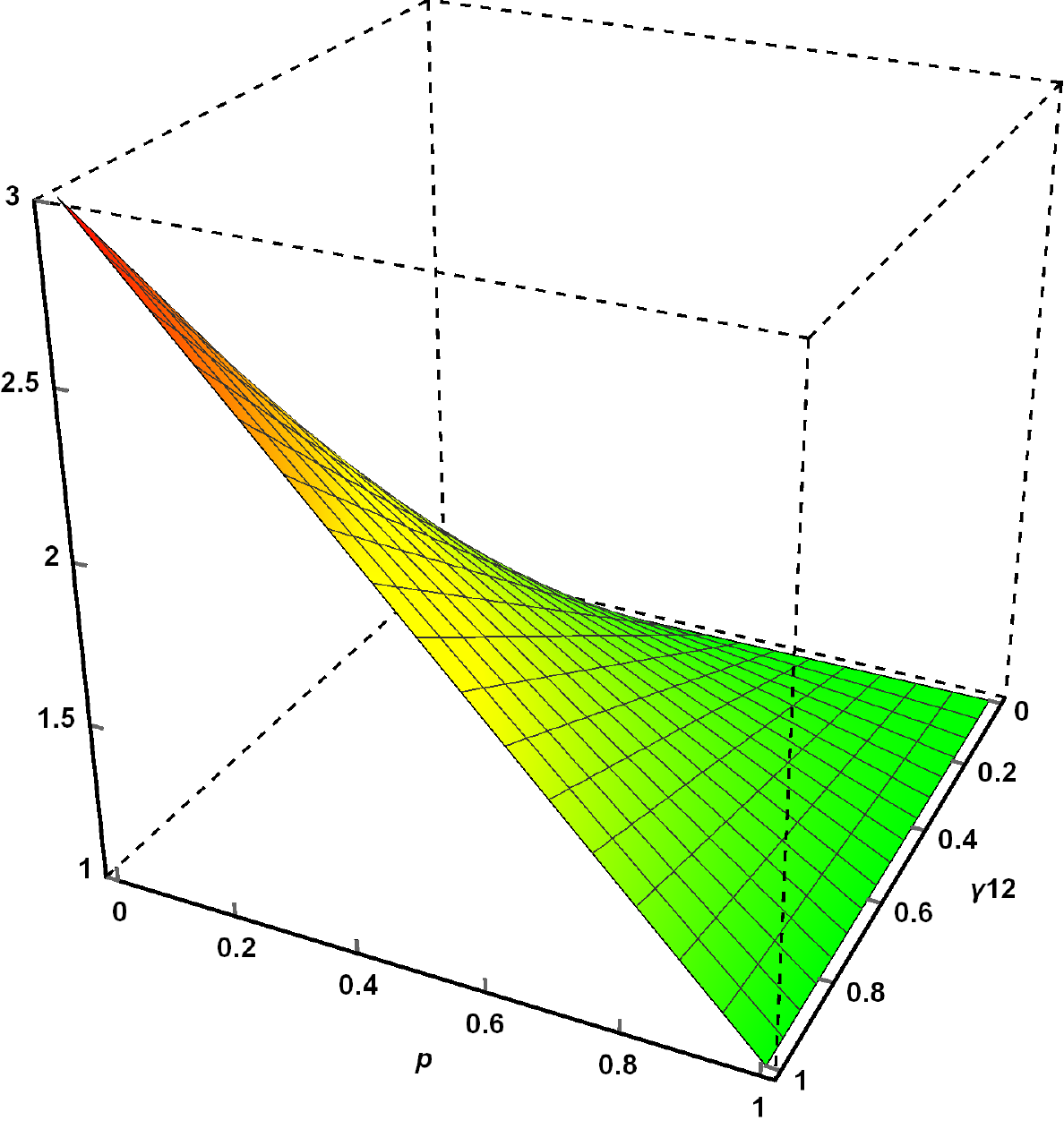}
 \centering
\caption{(color online): Quantum value of $(L_{\gamma})_Q$ is plotted against $p$ and $\gamma_{12}$. The relevant parameters are $\eta =1$, $\gamma_{23} = \gamma_{13} = 0$ and  $\theta =0.$}
\label{fig:1}
\end{figure}
It is found that quantum violation of standard LGI increases when channel parameter $\gamma_{12}$ increases but  $p$ decreases. The quantum value of  $(L_{\gamma})_Q$ reaches algebraic maximum $3$ for  $\gamma_{12} =1 $ and $p=0$.

\subsection{Quantum violation of third-order LGI}
The quantum mechanical expression of the third-order LGI (say, $(V_{\gamma})_Q$) in ineq.(\ref{vlgi1}) is calculated for the state  $|\psi(t_1)\rangle$ in Eq.(\ref{state}). The relevant three-time and one-time correlation functions  are given by
\begin{eqnarray}
\label{vg2}
\nonumber
&&(\langle M_{1} M_{2} M_3\rangle_{\gamma})_Q=\alpha ^3-\alpha ^2 \eta  (2 p-1) ((\gamma_{12}-1) \gamma_{23}-2 \gamma_{12})\\ \nonumber&&+\alpha  \eta ^2 (-2 \gamma_{12}+2 \gamma_{23} (\gamma_{12} (2 (p-1) p+1)-1)+3)\\ \nonumber&&+\eta  \cos (\theta ) \bigg[\alpha ^2 ((\gamma_{12}-1) \gamma_{23}-2 \gamma_{12}+3)-2 \alpha  \eta  (2 p-1) \\ \nonumber&& \times ((\gamma_{12}-1) \gamma_{23}-\gamma_{12})+\eta ^2 \big(\gamma_{23} \big(\gamma_{12} (1-2 p)^2-1\big)+1\big)\bigg]\\ &&+\gamma_{23} \eta ^3 (\gamma_{12}-2 \gamma_{12} p+2 p-1)
\end{eqnarray}
and
\begin{eqnarray}
\label{vg3}
(\langle M_{2} \rangle_{\gamma})_Q = \alpha -(\gamma_{12}-1) \eta  \cos (\theta )+\gamma_{12} \eta  (2 p-1)
\end{eqnarray}
respectively. For $\alpha=0$, using Eq.(\ref{Gm13}), Eq.(\ref{vg2}) and Eq.(\ref{vg3}) we have
\begin{eqnarray}
\nonumber
\label{gvsl1}
&&(V_{\gamma})_Q= \eta  \bigg[\gamma_{12}-\gamma_{13} \eta +\eta+(\gamma_{12}-1) \gamma_{23} \eta ^2 (-(2 p-1))\\ \nonumber &&-2 \gamma_{12} p +\cos (\theta ) \big(\gamma_{12}+\eta ^2 \left(\gamma_{23} \left(\gamma_{12} (1-2 p)^2-1\right)+1\right)\\  &&+\gamma_{13} \eta  (2 p-1)-1\big)\bigg]
\end{eqnarray}
If the system is evolved under GAD channel then the quantum violation of third-order LGI can be obtained for $\eta > 0.55$. The algebraic maximum $3$ of $(V_{\gamma})_Q$ is obtained for sharp measurement ($\eta=1$) at $\theta =0,\gamma_{12} = 1$, $\gamma_{23} = \gamma_{13} = 0$ and $p=0$.

For the case of $\alpha=1-\eta$, the quantum expression $(V_{\gamma})_Q$ is given in Eq.(\ref{gvsl2}). In this case similar to the case when standard LGI evolved under GAD channel, the condition for the quantum violation of third-order LGI is $\eta > 0.5$. However, maximum quantum violation is again obtained at $\eta = 1$ as expected.  The role of channel parameter responsible for the violation of third-order LGI can be seen from Fig. 2 where the quantum value of $(V_{\gamma})_Q$ is plotted against $p$ and $\gamma_{12}$.

\begin{figure}[ht]
\includegraphics[width=8 cm,height=7 cm]{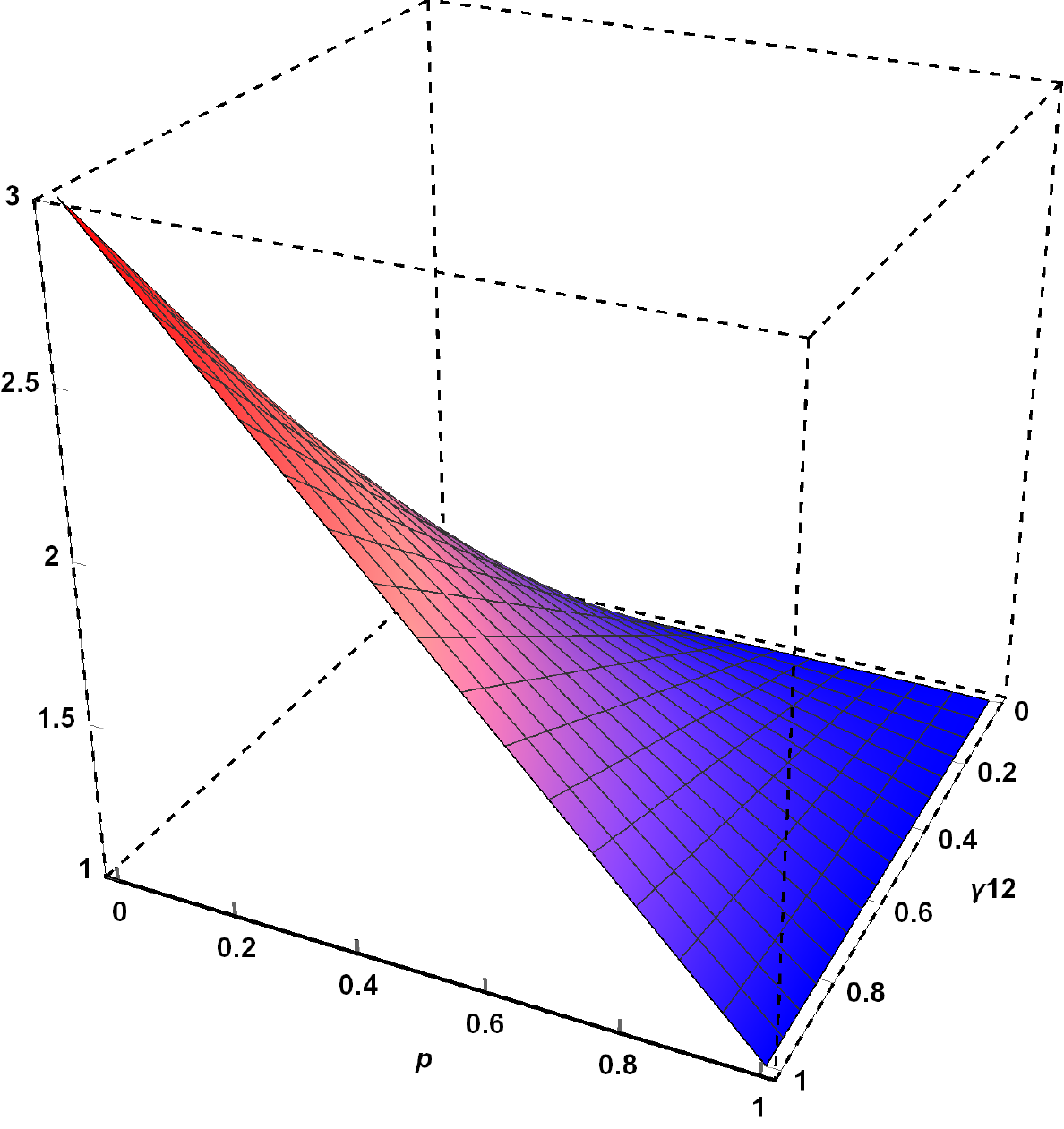}
\caption{(color online): The quantum value ${(V_{\gamma})}_Q$ of the third-order LGI expression is plotted against $p$ and $\gamma_{12}$. The relevant parameters are taken to be $\eta =1$, $\gamma_{23} = \gamma_{13} = 0$ and  $\theta =0 $.}
\label{fig:2}
\end{figure}

\section{NSIT conditions and violation of LGIs}
To critically examine the implications for the quantum violations of standard and third-order LGI upto algebraic maximum under non-unitary quantum channel, we define the degree of NSIT violations. Since the use of biased POVMs do not provide any further improvement of the quantum violations of LGIs, we restrict ourselves to sharp measurement ($\eta = 1$) case only.

The NSIT (AOT) condition assumes that the probability of obtaining an outcome of the measurements remains unaffected due to the prior (posterior) measurements. The conjunction of all the NSIT and AOT conditions ensures the existence of the global joint probability distribution $P(m_1, m_2,  m_3)$. The two-time NSIT conditions are given by
\begin{eqnarray}
NSIT_{(i)j} : P(m_j) = \sum_{m_1=\pm 1}P({m_i}, m_j)
\end{eqnarray}
Where, $i<j$. Similarly three-time NSIT conditions are given by
\begin{eqnarray}
\nonumber
NSIT_{1(2)3}:P(m_1, m_3)  \equiv \sum_{m_2 = \pm 1}P(m_1, m_2, m_3) \\
\end{eqnarray}  
and
\begin{eqnarray}
\nonumber
NSIT_{(1)23}:P(m_2, m_3) \equiv \sum_{m_1 = \pm 1}P(m_1, m_2, m_3)\\
\end{eqnarray}
The AOT conditions can be written as
\begin{eqnarray}
AOT_{i(j)}: P(m_i) \equiv \sum_{m_j = \pm 1}P(m_i, m_j)
\end{eqnarray}
Where, $i<j$, and
\begin{eqnarray}
AOT_{12(3)}: P(m_1, m_2)  \equiv \sum_{m_3 = \pm 1}P(m_1, m_2, m_3)
\end{eqnarray}
As already mentioned a suitable set of NSIT and AOT conditions provides the necessary and sufficient condition for macrorealism \cite{clemente15}. Then 
\begin{eqnarray}
\nonumber
NSIT_{(1)2} \wedge NSIT_{1(2)3} \wedge NSIT_{(1)23} \wedge AOT_{12(3)} \Leftrightarrow MR
\end{eqnarray}
However, NSIT conditions are necessary for LGIs and the violation of a LGI requires at least one of the NSIT condition to be violated but violation of one or all NSIT conditions do not warrant the violation of a given LGI. We can then write
\begin{eqnarray}
\nonumber
NSIT_{1(2)3} \wedge NSIT_{(1)23} \wedge AOT_{12(3)} \Rightarrow LGIs
\end{eqnarray}
The NSIT conditions can be shown to be violated within unitary quantum mechanics but AOT conditions are always satisfied. We first show here that how LG expressions can be cast in terms of degrees of violations of NSIT conditions. For this let us define the degree of violations of various NSIT conditions. The degree of violation for $NSIT_{(1)23}$ and $NSIT_{1(2)3}$ are quantified as
\begin{eqnarray}
\label{d11}
D_{(1)23}({m_2},{m_3}) = P({m_2},{m_3})-\sum_{m_1 = \pm 1} P({m_1},{m_2}, {m_3}) 
\end{eqnarray}
and
\begin{eqnarray}
\label{d22}
D_{1(2)3}(m_1 ,m_3) = P(m_1, m_3)-\sum_{m_2 = \pm 1} P(m_1, m_2, m_3)
\end{eqnarray}
respectively. Where, $D_{(1)23}(m_2,m_3)$ ($D_{1(2)3}(m_1,m_3)$) is the amount of disturbance (degree of violation of NSIT conditions) created by the measurement $M_1$($M_2$) at $t_1$ ($t_2$) to the measurements of $M_2$ and $M_3$ ($M_1$ and $M_3$) at $t_2$ and $t_3$ ($t_1$ and $t_3$) respectively.

 Note that $ D_{(1)23}(m_2, m_3)$ and $D_{1(2)3}(m_1, m_3)$ can take both positive and negative values.

In a macrorealist theory, the LGIs are derived by considering $D_{(1)23}(m_2, m_3)= D_{1(2)3}(m_1, m_3) = 0$. Since, AOT is assumed to be naturally satisfied. In QM it can be seen that the violation of standard LGI is achieved if either or both of $ D_{(1)23}(m_2, m_3)$ and $D_{1(2)3}(m_1, m_3)$ are non-zero (violation of NSIT condition). It can be understood through the study of the difference of $L$ and $L_{123}$, where the expression of $L_{123}$, obtained by considering all the three measurements for each correlations in $L$ given in Eq.(\ref{lgi1}), can be written as
\begin{eqnarray}
\label{lg123}
\nonumber
L_{123} &=&  -\langle M_{1} M_{2}\rangle_{123} + \langle M_{2} M_{3}\rangle_{123} +\langle M_{1} M_{3}\rangle_{123} \\ & = & 1 - 4 \beta
\end{eqnarray}
where, $\beta = P(M_{1}^{+} M_{2}^{+} M_{3}^{-})+P(M_{1}^{-} M_{2}^{-} M_{3}^{+})$,
\begin{eqnarray}
\nonumber
&& \langle M_{1} M_{2}\rangle_{123} = \sum_{m_1, m_2 = \pm 1} m_1 m_2 \sum_{m_3 = \pm 1} m_3 P(m_1, m_2, m_3) 
\end{eqnarray}
and similarly for $\langle M_{2} M_{3}\rangle_{123}$ and $\langle M_{1} M_{3}\rangle_{123}$. Now using Eq.(\ref{d11}-\ref{d22}), we can write 
\begin{eqnarray}
\nonumber
&&L - L_{123}\\ && =\sum_{m_2= m_3}D_{(1)23}(m_2, m_3)  -\sum_{m_2 \neq m_3}D_{(1)23}(m_2, m_3) \\ \nonumber&&-\sum_{m_1 \neq m_3}D_{1(2)3}(m_1, m_3)+\sum_{m_1 = m_3}D_{1(2)3}(m_1, m_3)
\end{eqnarray}
Since $L \leq 1$ and $L_{123} = 1- 4 \beta $ we obtain
\begin{eqnarray}
\label{NLgi1}
&&\sum_{m_2 = m_3}D_{(1)23}(m_2, m_3)+\sum_{m_1 = m_3}D_{1(2)3}(m_1, m_3) \leq 2 \beta
\end{eqnarray}
We have thus written the standard LG expression in terms of the degrees of NSIT violations. Hence, for the violation of LGI, one needs
\begin{eqnarray}
\label{Lgiv1}
&&\sum_{m_2 = m_3}D_{(1)23}(m_2, m_3)+\sum_{m_1 = m_3}D_{1(2)3}(m_1, m_3)> 2 \beta
\end{eqnarray}
should be satisfied. Hence, we can say that the violation of standard LGI depends on interplay between the violations of NSITs conditions. 
\begin{figure}[ht]
\includegraphics[width=1\linewidth]{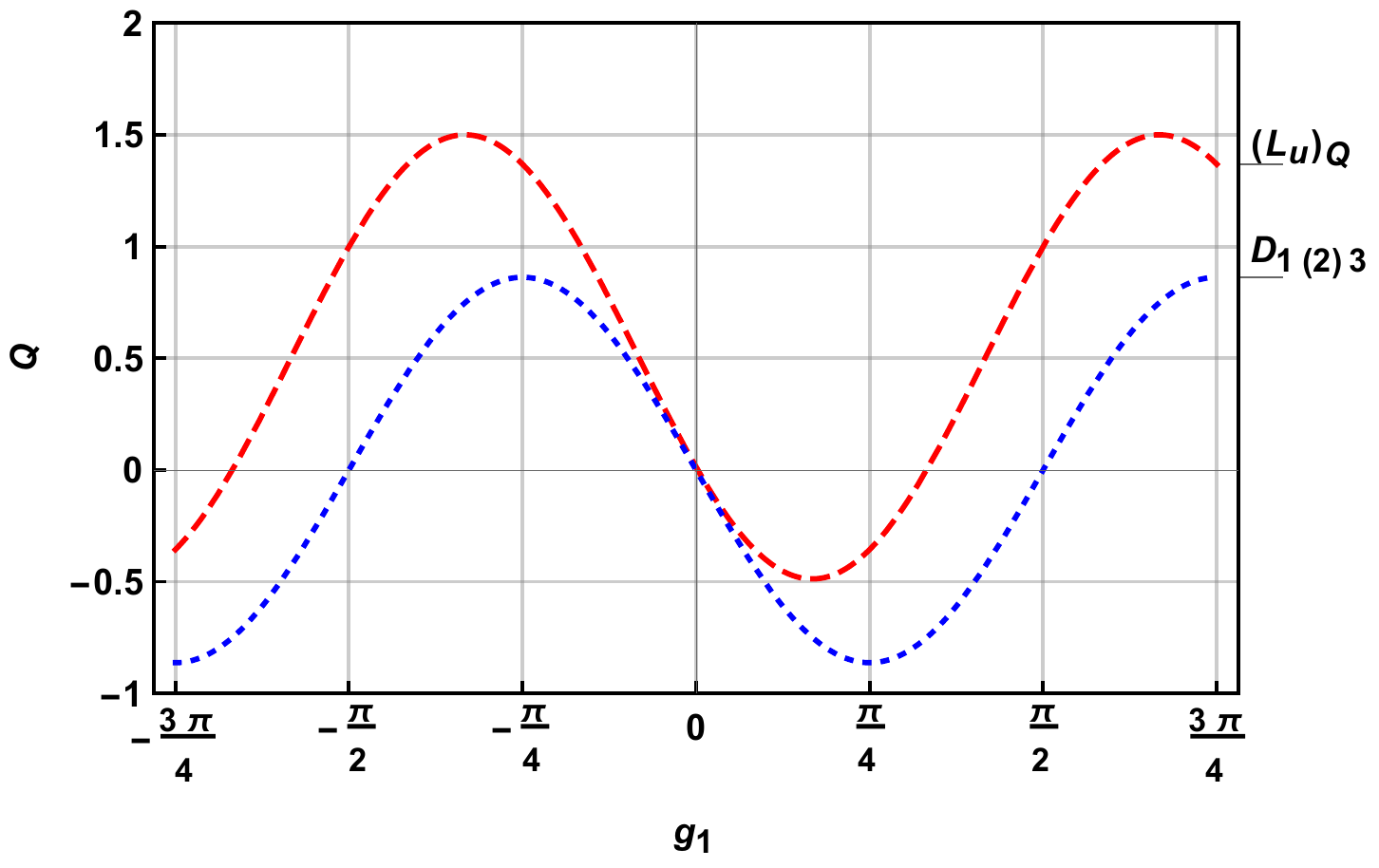}
\caption{(color online): Quantum mechanical expression $(L_{u})_Q$ in Eq.(\ref{sl1}) (Dashed Red) and  $D_{1(2)3}$ in Eq.(\ref{ud11}) (Dotted Blue) are plotted against $g_1$ at $g_2 = \pi/6$ .}
\end{figure}
\begin{figure}[ht]
\includegraphics[width=1\linewidth]{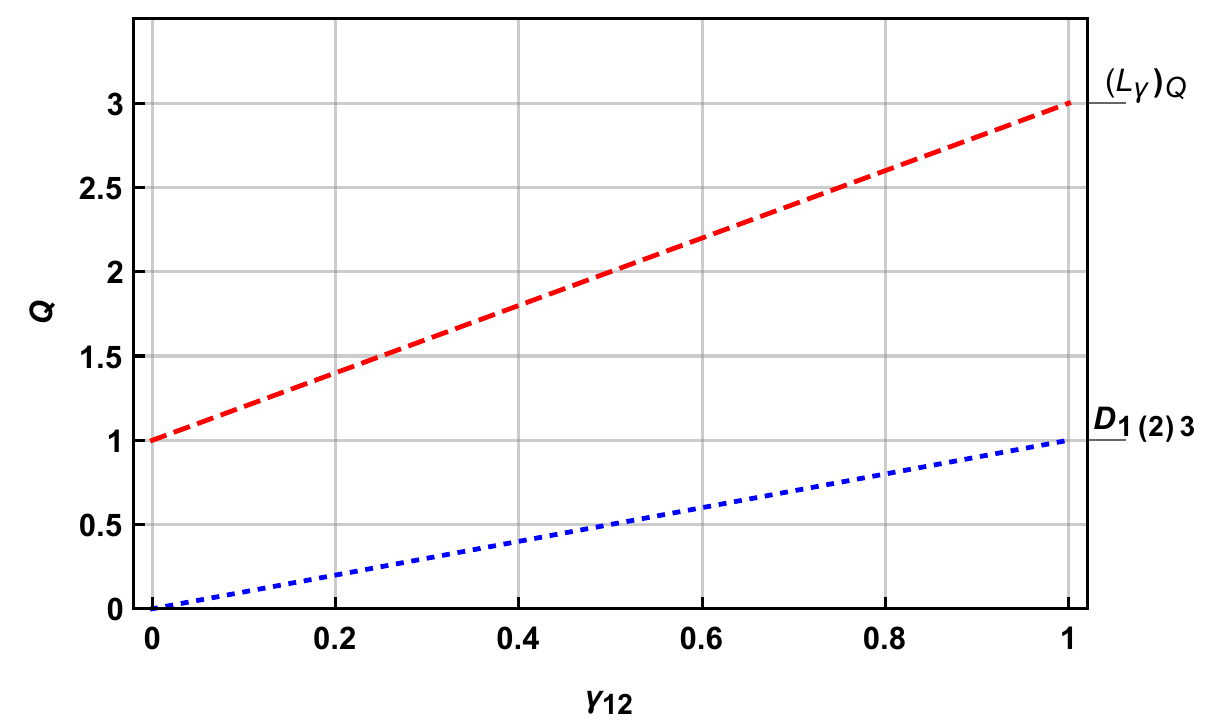}
\caption{(color online): Quantum mechanical expression $(L_{\gamma})_Q$  in Eq.(\ref{gsl1}) (Dashed Red) and $D_{1(2)3}$  in Eq.(\ref{gd11}) (Dotted Blue) are plotted against $\gamma_{12}$ at $P=0, \theta =0$ and $\gamma_{23} = \gamma_{13} = 0$.}
\label{fig:4}
\end{figure}
\begin{figure}[ht]
\includegraphics[width=1\linewidth]{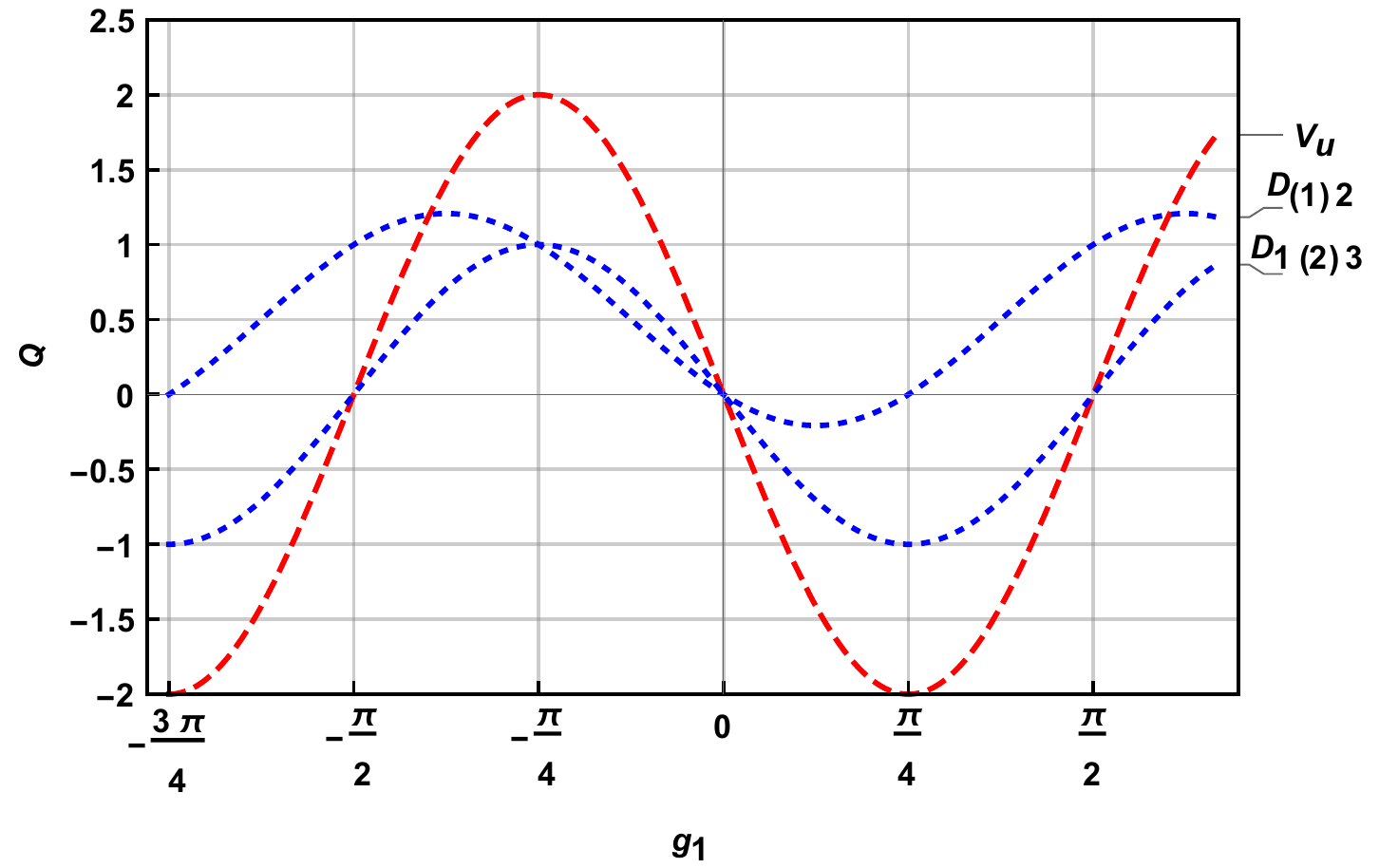}
\caption{(color online): Quantum mechanical expression $(V_u)_Q$ in Eq.(\ref{vsl2}) (Dashed Red), $D_{1(2)3}$ and $D_{(1)2}$ (Dotted Blue) are plotted against $g_1$ at $g_2 = \pi/4$}
\label{fig:5}
\end{figure}
\begin{figure}[ht]
\includegraphics[width=1\linewidth]{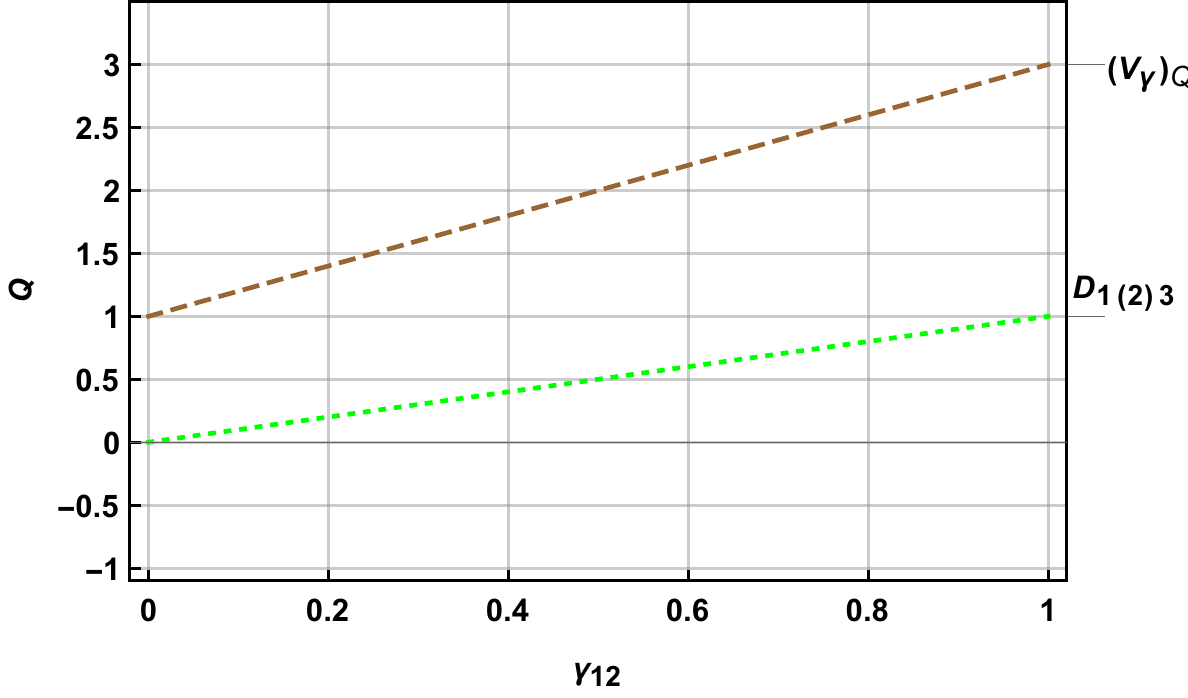}
\caption{(color online): Quantum mechanical expression $(V_{\gamma})_Q$ in Eq.(\ref{gvsl1}) (Dashed Brown) and  $D_{1(2)3}$ in Eq.(\ref{gd11}) (Dotted Green) are plotted against $\gamma_{12}$ at $P=0, \theta =0$ and $\gamma_{23} = \gamma_{13} = 0$.}
\label{fig:6}
\end{figure}
Next, we consider the third-order LGI in Eq.(\ref{vlgi1}). In order to cast the third-order LGI in terms of degree of violation of NSIT conditions by following the step earlier we consider the difference between $V$ and $V_{123}$. Here the expression of $V_{123}$, obtained by considering the measurements of all the three observables for each correlations in $V$ given in Eq.(\ref{vlgi1}). Thus $V_{123}$ can be written as
\begin{eqnarray}
\label{vlg123}
\nonumber
V_{123} &=&  \langle M_{1} M_{2} M_{3}\rangle_{123} + \langle M_{1} M_{3}\rangle_{123} -\langle M_{2}\rangle_{123} \\ & = & 1 - 4 \delta
\end{eqnarray}
where, $\delta = P(M_{1}^{-} M_{2}^{+} M_{3}^{+})+P(M_{1}^{+} M_{2}^{+} M_{3}^{-})$ and
\begin{eqnarray}
\nonumber
&&\langle M_{1} M_{2} M_{3}\rangle_{123} = \sum_{m_1,m_2, m_3 = \pm 1} m_1 m_2 m_3 P(m_1, m_2, m_3), \\ \nonumber && \langle M_{2}\rangle_{123} =  \sum_{m_2  = \pm 1} m_2 \sum_{m_1, m_3 = \pm 1} m_1 m_3 P(m_1, m_2, m_3)
\end{eqnarray}
 Since $\langle M_{1} M_{2} M_{3}\rangle_{123} = \langle M_{1} M_{2} M_{3}\rangle$, then disturbance comes due to $\langle M_{1} M_{3}\rangle$ and $\langle M_{2}\rangle$ (amount of disturbance created by $M_1$ at time $t_1$). Then the  relevant three-time and two-time NSIT conditions are quantified as $D_{1(2)3}(m_1 ,m_3)$ in Eq.(\ref{d22}) and 
\begin{eqnarray}
\label{dm2}
D_{(1)2}(m_2)&=&P(m_2)-\sum_{m_1} P(m_1, m_2)
\end{eqnarray}
respectively. Using Eq.(\ref{d22}) and NSIT condition in Eq.(\ref{dm2}) we can write
\begin{eqnarray}
\label{vv123}
V - V_{123} &=& 2\sum_{m_1= m_3}D_{1(2)3}(m_1, m_3)-D_{(1)2}(m_2)
\end{eqnarray}
By noting $V \leq 1$ and substituting $V_{123} = 1 - 4 \delta$ from Eq.(\ref{vlg123}), we obtain
\begin{eqnarray}
 2\sum_{m_1= m_3}D_{1(2)3}(m_1, m_3)-D_{(1)2}(m_2) \leq 4 \delta
\end{eqnarray}
Then for the violation of third-order LGI either 
\begin{eqnarray}
\label{v1}
 2\sum_{m_1= m_3}D_{1(2)3}(m_1, m_3)-D_{(1)2}(m_2) \geq 4 \delta
\end{eqnarray}
From Eq.(\ref{v1})  we can say that the violation of third-order LGI depends on the interplay between degree of violation of NSIT conditions. For the violation of third-order LGI the threshold value $\delta$ has to be reached. Henceforth, for avoiding the clumsiness of the notation in disturbance in LGIs, we denote $ D_{(1)23}(m_2, m_3),  D_{1(2)3}(m_1, m_3)$ and $D_{(1)2}(m_2)$ by $D_{(1)23}$, $D_{1(2)3}$ and  $D_{(1)2}$ respectively.

The implications of quantum violation of standard LGI (Eq.(\ref{lgi1})) can be explicitly examined by calculating degree of violation of  $NSIT_{(1)23}$ and $NSIT_{1(2)3}$ conditions.  For the quantum state given in Eq.(\ref{state}) evolved under unitary dynamics, the quantum expression of disturbance quantifying the degree of violation of  $NSIT_{(1)23}$ and $NSIT_{1(2)3}$  conditions are obtained as,
\begin{eqnarray}
D_{(1)23} = 0    
\end{eqnarray}
and
\begin{eqnarray}
\label{ud11} 
   D_{1(2)3} =  \sin (2 g_1) \sin (2 g_2) 
\end{eqnarray}
respectively.
However, for the same quantum state and sharp measurement, if evolution is followed by GAD channel, the quantum expression of disturbance for  $NSIT_{1(2)3}$,  $NSIT_{(1)23}$ and $D_{1(2)3}$ are obtained as
\begin{eqnarray}
&&D_{(1)23} = 0;\\
\label{gd11}
\nonumber
&& D_{1(2)3} =   \frac{1}{2}  (\gamma_{12} (\gamma_{23}-1)+\gamma_{13}-\gamma_{23}) [-1 -\cos (\theta ) (1 -2  p)] \\
\end{eqnarray}
respectively. Since, in both case of quantum system evolved under unitary dynamics and GAD channel $D_{(1)23}= 0$ we can say that the degree of violation of $NSIT_{1(2)3}$ (non-zero value of $D_{1(2)3}$) is responsible for the quantum violation of standard LGI. In Fig.3   we have plotted the quantum expression of standard LGI $(L_{u})_Q$ in Eq.(\ref{sl1}) and $D_{1(2)3}$ in Eq.(\ref{ud11}). The quantum expression of $(L_{\gamma})_Q$ in Eq.(\ref{gsl1}) and $D_{1(2)3}$ in Eq.(\ref{gd11}) is plotted in Fig.4. 

 We further study the degree of violation of NSITs conditions for the case of third-order LGI. The quantum expression of disturbance quantifying degree of violation of  $NSIT_{(1)23}$ and $NSIT_{(1)2}$ responsible for the violation third-order LGI in Eq.(\ref{vlgi1}) for the quantum state given in Eq.(\ref{state}) evolved under unitary dynamics and POVMs in Eq.(\ref{GO}) for $\eta = 1$ are $D_{1(2)3} =  \sin (2 g_1) \sin (2 g_2)$ given in Eq.(\ref{ud11}) and
\begin{eqnarray}
\label{ud1111}
\nonumber
&&D_{(1)2}=\sin (g_1) \bigg[2 \sin (g_1) \cos ^2(\theta)-\cos (g_1) \sin (2 \theta) \sin (\phi )\bigg] \\
\end{eqnarray}
 respectively. The quantum expression of $(V_{u})_Q$ in Eq.(\ref{vsl2}), $D_{1(2)3}$ in Eq.(\ref{ud11})   and $D_{(1)2}$ in Eq.(\ref{ud1111}) is plotted in Fig.5 to demonstrate the variation of quantum violation with the violation of $NSIT_{(1)2}$.

However, if system evolves under GAD channel, NSIT conditions $NSIT_{1(2)3}$ and $NSIT_{(1)2}$ are responsible for the violation of third-order LGI. The quantum expression of the degree of violation of $NSIT_{1(2)3}$ and $NSIT_{(1)2}$ are $D_{1(2)3}$ given by Eq.(\ref{gd11}) and $ D_{(1)2}=0$. The quantum expression of third-order LGI under non-unitary quantum channel $(V_{\gamma})_Q$ and  $D_{1(2)3}$ in Eq.(\ref{gd11}) are plotted against $\gamma_{12}$ in Fig.6. It is seen that like standard LGI non-zero value of $D_{1(2)3}$ is responsible for the quantum violation of third-order LGI.
\section{Discussion}
In this paper, we have studied two formulations of three-time LGIs, viz., the standard and third-order LGI  when the system evolves under  non-unitary quantum channel  between two measurements. We have considered simple GAD channel for our demonstration and showed that the respective L$\ddot{u}$ders bounds of both standard and third-order LGIs are violated. Interestingly, quantum violations of both the forms of LGIs reach their algebraic maximum $3$ for sharp measurement. We have also examined the quantum violations of standard and third-order LGIs for the cases when the measurements are characterized by one-parameter unbiased POVMs and two-parameter biased POVMs. It is found that when POVMs are unbiased and system evolves under GAD channel, the quantum violations of both standard and third-order LGI can be obtained for lower value of the sharpness parameter compared to the case when system evolves under unitary dynamics.  We may remark here that although our calculation is based on a qubit system but the feature is generic and valid for dichotomic measurements in ay arbitrary dimensional system.

As already discussed earlier, the LG test performed through three successive measurements corresponds to a  stronger reading of non-invasive measurability, and consequently stronger notion of macrorealism \cite{halli16}. In such a case, no set of LGIs provide the necessary and sufficient condition for the stronger notion of macrorealism, i. e., the violation of a LGI warrants the violation of macrorealism but converse is not true. However, a suitably formulated set of NSIT conditions provide the necessary and sufficient condition \cite{clemente15}. This implies that the quantum violation of a given LGI inevitably requires one of the NSIT conditions to be violated. We critically examined the relation between the violation of standard and third-order LGI with NSIT conditions. By considering the degree of violations of various NSIT conditions we showed that how a given LGI can be cast in terms of them. This enables us to argue that mere the violation of  NSIT conditions do not warrant the violation of standard or third-order LGIs, the interplay between the violations of various NSIT conditions and some threshold values play an important role.

\begin{widetext}
\appendix
\section{}
If the state $|\psi(t_1)\rangle$ given in Eq.(\ref{state}) evolved under unitary dynamics, quantum expression of joint expectation values in LHS of ineq.(\ref{lgi1}) for the  POVMs in Eq.(\ref{GO}) are given by 
\begin{eqnarray}
\label{m12}
(\langle M_{1} M_{2}\rangle_u)_Q &=&\alpha ^2+\eta ^2 \cos (2 g_1)+\eta  \cos (g_1) \bigg[\big(\chi_1-\chi_2\big) \sin (g_1) \sin (2 \theta) \sin (\phi )2 \alpha  \cos (g_1) \cos (2 \theta)\bigg]
\end{eqnarray}
\begin{eqnarray}
\label{m23}
 (\langle M_{2} M_{3}\rangle_u)_Q &=&\frac{1}{2} \bigg[2 \alpha ^2+8 \alpha  \cos ^2(g_2) \sin (\theta) \cos (\theta) \sin (\phi ) +\cos (2 g_1)4 \alpha  \cos ^2(g_2) \cos (2 \theta)+2 \eta  \cos (2 g_2)\\ \nonumber&+&\eta \sin (2 g_2) \bigg(\big(\chi_2 - \chi_1\big)\big(\sin (2 g_1)   \cos (2 \theta)-\cos (2 g_1) \sin (2 \theta) \sin (\phi )\big)\bigg)\bigg]
\end{eqnarray}
and
\begin{eqnarray}
\label{m13}
(\langle M_{1} M_{3}\rangle_u)_Q &=& \alpha ^2+\eta ^2 \cos (2 (g_1+g_2))+\eta  \cos (g_1+g_2) \bigg[\big(\chi_1 -\chi_2\big) \sin (2 \theta) \sin (\phi ) \sin (g_1+g_2)\\ \nonumber&+&2 \alpha  \cos (2 \theta) \cos (g_1+g_2)\bigg]
\end{eqnarray}
respectively, where, $\chi_1=\sqrt{(1+\alpha)^2 -\eta^2 }$ and $\chi_2= \sqrt{(1-\alpha)^2 -\eta^2 }$. Using Eq.(\ref{m12}),  Eq.(\ref{m23}) and  Eq.(\ref{m13}) for $\alpha=1-\eta$, LHS of standard LGI in ineq.(\ref{lgi1}) reduces to
\begin{eqnarray}
\label{bsl1}
(L_{u})_{Q}&&= \frac{1}{2} \bigg[2 (\eta -1)^2+\eta  \bigg(\sin (2 \theta) \sin (\phi ) \left(4 \sqrt{1-\eta } \sin (g_2) \cos (2 g_1+g_2)-4 (\eta -1) \sin (2 g_1) \cos ^2(g_2)\right)\\ \nonumber&&+\cos (2 \theta) \left(4 (\eta -1) \sin (g_2) \sin (2 g_1+g_2)-2 \sqrt{1-\eta } \sin (2 g_1) \sin (2 g_2)\right)+4 \eta  \cos (g_1) \cos (g_1+2 g_2)\\ \nonumber&&+\cos (2 g_1) \left(-2 \eta +2 \sqrt{1-\eta } \sin (2 g_2) \sin (2 \theta) \sin (\phi )-4 (\eta -1) \cos ^2(g_2) \cos (2 \theta)\right)\bigg)\bigg]
\end{eqnarray}
In case of third-order LGI the quantum expression of  $\langle M_{1} M_{2} M_3\rangle $ and $\langle M_2 \rangle$ in LHS of ineq.(\ref{vlgi1}) are obtained as
\begin{eqnarray}
\label{m123}
(\langle M_{1} M_{2} M_3\rangle_u)_Q &&= e^{-i \phi } \bigg[\cos ^2(g_1) \big[\alpha  e^{i \phi } (\alpha ^2+\eta ^2)+2 \alpha ^2 \eta  e^{i \phi } \cos (2 \theta)+\eta  e^{i \phi } \cos (2 g_2) (\alpha ^2+\eta ^2) \cos (2 \theta)\\ \nonumber &&+i \eta \big(1+\alpha ^2-\eta ^2-\chi_3 \big)   \sin (g_2) \cos (g_2) \sin (\theta) \cos (\theta) (1-e^{2 i \phi })  \big]\\ \nonumber &&+e^{i \phi } \sin ^2(g_1) \big[\alpha  (\alpha^2 -\eta^2 )-2 \eta  \big(-\chi_3+1+\alpha ^2-\eta ^2\big) \sin (g_2) \cos (g_2) \sin (\theta) \cos (\theta) \sin (\phi )\\ \nonumber &&
-\eta (\alpha^2 - \eta^2 ) \cos (2 g_2) \cos (2 \theta)\big]+\eta  e^{i \phi }\sin (2 g_1) \cos (g_2) \big(\sin (g_2) (\alpha  \cos (2 \theta)+\eta )\\ \nonumber &&-\alpha  \cos (g_2) \sin (2 \theta) \sin (\phi )\big)  \big(\chi_2-\chi_1 \big)\bigg]
\end{eqnarray}
and
\begin{eqnarray}
\label{m2}
(\langle M_{2}\rangle_u)_Q && = \alpha +\eta  \sin (2 g_1) \sin (2 \theta ) \sin (\phi )+\eta  \cos (2 g_1) \cos (2 \theta )
\end{eqnarray}
respectively, where $\chi_3 =\sqrt{1-(\alpha - \eta)^2}\sqrt{1-(\alpha + \eta)^2} $. The quantum expression of third-order LGI is obtained by substituting Eq.(\ref{m13}),  Eq.(\ref{m123}) and Eq.(\ref{m2}) in LHS of ineq.(\ref{vlgi1}) for $\alpha = 1-\eta$ is given by
\begin{eqnarray}
\label{bvl1}
(V_{u})_{Q}&&=\eta  \bigg[\cos (2 \theta) \big((\eta -1) \left(\eta +\left(\sqrt{1-\eta }+1\right) \sin (2 g_1) \sin (2 g_2)-2\right)+(\eta -2) \cos (2 g_1) (\eta +(\eta -1) \cos (2 g_2))\\ \nonumber&&+\eta ^2 \cos (2 g_2)\big)+\sin ^2(g_1) (2 \eta +(\eta -1) \sin (2 g_2) \sin (2 \theta) \sin (\phi )-3)+\sqrt{1-\eta } \sin (2 \theta) \sin (\phi ) \sin (2 (g_1+g_2))\\ \nonumber&&-2 \sqrt{1-\eta } \eta  \sin (2 g_1) \cos ^2(g_2) \sin (2 \theta) \sin (\phi )+\sin (2 g_1) \sin (2 \theta) \sin (\phi ) \left(2 \sqrt{1-\eta } \cos ^2(g_2)-1\right)+\eta  \cos (2 (g_1+g_2))\bigg]\\ \nonumber&& -(\eta -1) \cos ^2(g_1) \left(2 (\eta -1) \eta +2 \eta ^2 \cos (2 g_2)+\eta  \sin (2 g_2) \sin (2 \theta) \sin (\phi )+1\right)+ (\eta -1) \eta\\ \nonumber&&-4 \sqrt{1-\eta } \eta ^2 \sin (g_1) \cos (g_1) \sin (g_2) \cos (g_2)+\sin ^2(g_1)
\end{eqnarray}
Next, if the state $|\psi(t_1)\rangle$ given in Eq.(\ref{state}) evolved under GAD channel, the quantum expression of standard LGI and third-order LGI for the  POVMs in Eq.(\ref{GO}) with $\alpha=1-\eta$ are obtained as 
\begin{eqnarray}
\label{gsl2}
(L_{\gamma})_Q&=& (\eta -1)^2+\eta ^2 \left(\gamma_{12}-\gamma_{13}-\gamma_{23}+\gamma_{12} \gamma_{23} (1-2 p)^2+1\right)-\eta  \cos (\theta ) (\gamma_{12} (-\gamma_{23})+\gamma_{12}+\gamma_{13}+\gamma_{23}\\ \nonumber&+&2 \eta  (\gamma_{12} (\gamma_{23} p+p-1)-p (\gamma_{13}+\gamma_{23})+1)-2)+(1-\eta ) \eta  (2 p-1) (\gamma_{12} (-\gamma_{23})+\gamma_{12}+\gamma_{13}+\gamma_{23})
\end{eqnarray}
and
\begin{eqnarray}
\label{gvsl2}
(V_{\gamma})_Q &=&1-4 \eta ^3 (\gamma_{12} p-1) (\gamma_{23} p-1)+ \eta  \cos (\theta ) \bigg[\gamma_{12} \gamma_{23}-\gamma_{12}-\gamma_{13}-\gamma_{23}+2 \eta  (\gamma_{12}-2 (\gamma_{12}-1) \gamma_{23} p\\ \nonumber&+&2 \gamma_{12} p+\gamma_{13} p-4)+4 \eta ^2 (\gamma_{12} p-1) (\gamma_{23} p-1)+4\bigg]+2 \eta ^2 (\gamma_{12}+2 \gamma_{23} p (\gamma_{12} p-1)-4 \gamma_{12} p-\gamma_{13} p+4)\\ \nonumber&-&\eta  (\gamma_{13}+\gamma_{23}+\gamma_{12} (\gamma_{23}-1) (2 p-1)-2 p (\gamma_{13}+\gamma_{23})+4)
\end{eqnarray}
respectively.

\end{widetext}


\begin{thebibliography}{99}

\bibitem{sch} E. Schroedinger, Naturwissenschaften, \textbf{23}, 807 (1935).
\bibitem{zur} H. D. Zeh,  Found. Phys. 1, \textbf{69} (1970), W. H. Zurek, Phys. Rev. D \textbf{26}, 1862 (1982).
\bibitem{bruk} J. Kofler and C. Brukner, Phys. Rev. Lett. \textbf{99}, 180403 (2007).
\bibitem{ghi} G. C. Ghirardi, A. Rimini, and T. Weber, Phys. Rev. D \textbf{34}, 470 (1986).
\bibitem{bell64}J. S. Bell, Physics \textbf{1}, 195 (1964); J. F. Clauser, M. A. Horne, A. Shimony and R. A. Holt, 	Phys. Rev. Lett. \textbf{23}, 880, (1969).
\bibitem{lg85} A. J. Leggett and A. Garg, Phys. Rev. Lett. \textbf{54}, 857 (1985).
\bibitem{lg02} A. J. Leggett, J. Phys.: Condens. Matter \textbf{14}, R415 (2002).
\bibitem{pan18} A. K. Pan, Md. Qutubuddin and S. Kumari, Phys. Rev. A, \textbf{98}, 062115(2018).
\bibitem{halli19} J. J. Halliwell,Phys. Rev. A \textbf{99}, 022119 (2019). 
\bibitem{halli20} J. J. Halliwell and C. Mawby, Phys Rev A 102,,012209 (2020).
\bibitem{kofler13} J. Kofler and C. Brukner, Phys. Rev. A \textbf{87}, 052115 (2013).
\bibitem{clemente15} L. Clemente and J. Kofler, Phys. Rev. A, \textbf{91}, 062103 (2015).

\bibitem{maroney14} O. J. E. Maroney and C. G Timpson, arxiv: 1412.613v1.

\bibitem{emary12} C. Emary, N. Lambert, and F. Nori, Phys. Rev. B \textbf{86}, 235447
(2012).
\bibitem{budroni15} C. Budroni \emph{et al}.,  Phys. Rev. Lett. \textbf{115}, 200403 (2015).
\bibitem{saha15} D. Saha, S. Mal, P. K. Panigrahi, and D. Home, Phys. Rev. A \textbf{91}, 032117 (2015).

\bibitem{halli16} J. J. Halliwell, Phys. Rev. A, \textbf{93}, 022123 (2016).


\bibitem{budroni14} C. Budroni and C. Emary,  Phys. Rev. Lett. \textbf{113}, 050401 (2014).
\bibitem{swati17} S. Kumari and A. K. Pan, Euro. Phys. Lett. \textbf{118}, 50002 (2017).
\bibitem{pan17} S. Kumari and A. K. Pan, Phys. Rev. A \textbf{96}, 042107 (2017).
\bibitem{hal19a} S-S. Majidy, H. Katiyar, G. Anikeeva, J. J. Halliwell and R. Laflamme, Phys. Rev. A \text{100}, 042325 (2019).
\bibitem{hal21} S-S. Majidy, J.J. Halliwell, R. Laflamme, Phys. Rev. A \textbf{103}, 062212 (2021).
\bibitem{hal211} J. J. Halliwell, A. Bhatnagar, E. Ireland, H. Nadeem, V. Wimalaweera, Phys. Rev. A \textbf{103}, 032218 (2021).
\bibitem{pan20} A. K. Pan,  Phys. Rev. A \textbf{102}, 032206 (2020).
\bibitem{sk} S. Kumari and A. K. Pan, J. Phys. A: Math. Theor.\textbf{ 54} , 035301(2021).
\bibitem{kofler08} J. Kofler and C. Brukner, Phys. Rev. Lett. \textbf{101}, 090403 (2008).
\bibitem{emary} C. Emary, N. Lambert and F. Nori,  Rep. Prog. Phys. \textbf{77}, 016001 (2014).

\bibitem{lambert} N. Lambert, K. Debnath, A. F. Kockum, G. C. Knee, W. J. Munro, and F. Nori,  Phys. Rev. A \textbf{94}, 012105 (2016).

\bibitem{goggin11} M. E. Goggin \emph{et al.}  Proc. Natl. Acad. Sci. U.S.A. \textbf{108}, 1256 (2011).
\bibitem{knee12} G. C. Knee \emph{et al}., Nat. Commun. \textbf{3}, 606 (2012).
\bibitem{laloy10} A. Palacios-Laloy \emph{et al}., Nat. Phys. \textbf{6}, 442 (2010).
\bibitem{george13} R. E. George \emph{et al}., Proc. Natl. Acad. Sci. U.S.A. \textbf{110}, 3777 (2013).


\bibitem{knee16} G. C. Knee \emph{et al}., Nat. Commun. \textbf{7}, 13253 (2016).
\bibitem{kati1} H. Katiyar, A. Brodutch, D. Lu and R. Laflamme, N. J. Phys. \textbf{19}, 023033 (2017).
\bibitem{wang02} K. Wang, C. Emary, X. Zhan, Z. Bian, J. Li and P. Xue, Opt. Exp. \textbf{25}, 31462 (2017).

\bibitem{akumari18} A. Kumari, Md. Qutubuddin and  A. K. Pan,  Phys. Rev. A \textbf{98},  042135 (2018).



\bibitem{halli17} J. J. Halliwell, Phys. Rev. A \textbf{96}, 012121 (2017).


\bibitem{halli19a} J. J. Halliwell and C. Mawby, Phys. Rev. A \textbf{100}, 042103 (2019).
\bibitem{wild} M. Wild and A. Mizel Found. Phys. \textbf{42}, 256 (2012).

\bibitem{clemente16} L. Clemente and J. Kofler, Phys. Rev. Lett. \textbf{116}, 150401 (2016).



\bibitem{bender02} C. M. Bender, D. C. Brody, and H. F. Jones, Phys. Rev. Lett. \textbf{89},
270401 (2002).


\bibitem{Karthik} H S Karthik, A. S. Hejamadi, and A. R. Usha Devi, Phys. Rev. A, \textbf{103}, 032420 (2021).
\bibitem{varma01} A. V. Varma and  S. Das, arXiv:2012.13415.
\bibitem{varma} A. V. Varma, I. Mohanty and  S. Das,  J. Phys. A: Math. Theor. \textbf{54}, 115301 (2021).
\bibitem{javid20} J. Naikoo, S. Kumari, A. K Pan, S. Banerjee, J. Phys. A: Math. Theor.\textbf{ 54}, 275303 (2021).
\bibitem{niel} M. Nielsen and I. Chuang, \textit{Quantum Computation and Quantum Information}
(Cambridge University Press, Cambridge, 2000).

\bibitem{sri} R. Srikanth and S. Banerjee, Phys. Rev. A \textbf{77}, 012318 (2008).
\bibitem{budroni13} C. Budroni, T. Moroder, M. Kleinmann, and O. G$\ddot{u}$hne,  Phys. Rev. Lett. \textbf{111}, 020403 (2013).


\end{thebibliography}
\end{document}